# Theoretical development on the isosteric heat of adsorption and experimental confirmation

Kyaw Thu [1,2,*], Sourav Mitra [3] and Bidyut Baran Saha [3,4,*]

[1] Kyushu University Program for Leading Graduate School, Green Asia Education Center,
Interdisciplinary Graduate School of Engineering Sciences,
Kyushu University, Kasuga-koen 6-1, Kasuga-shi, Fukuoka 816-8580, Japan
[2] Interdisciplinary Graduate School of Engineering Sciences
Kyushu University, Kasuga-koen 6-1, Kasuga-shi, Fukuoka 816-8580, Japan
[3] International Institute for Carbon-Neutral Energy Research (WPI-I2CNER),
Kyushu University, 744 Motooka, Nishi-ku, Fukuoka 819-0395, Japan
[4] Mechanical Engineering Department, Kyushu University,
744 Motooka, Nishi-ku, Fukuoka 819-0395, Japan

*Corresponding Authors' email: kyaw.thu.813@m.kyushu-u.ac.jp (KT),
saha.baran.bidyut.213@m.kyushu-u.ac.jp (BBS)

**ABSTRACT**

Theoretical framework for the isosteric heat of adsorption is developed treating the effects of the non-ideal gas behavior and the adsorbed phase volume. Rigorous thermodynamic treatment for the adsorbed phase volume is presented for multi-layer adsorption from low to high pressures. The proposed model for the isosteric heat of adsorption along with the adsorbed phase volume is validated and verified using experimental data for several, judiciously selected adsorbent + adsorbate (nonpolar molecules) systems. The predictions by the model exhibit excellent agreement with the experimental data for both adsorption isotherms and the isosteric heat of adsorption.

In adsorption science, isosteric heat of adsorption is a crucial thermodynamic quantity in auditing the energy associated with the adsorption systems. Accurate knowledge on the isosteric heat involved during sorption (adsorption and desorption) processes leads to an excellent system design and cost benefits[1,2]. Various adsorbent + adsorbate pairs have been successfully applied in broad range of industrial and commercial applications starting from separation for environmental control[3], energy storage[4,5], HVAC (cooling, refrigeration and dehumidification)[6–13] and desalination[14–17]. Advancements in development of the tailored adsorbents for a typical adsorbate further calls for the precise estimation of the energy involved at a particular amount of adsorption.

In general, the heat of adsorption can be viewed as the release of the kinetic energy from the molecules of the bulk phase being adsorbed onto the surfaces of the adsorbent. In scientific terms, the isosteric heat of adsorption is the difference between the molar enthalpy of the gaseous phase and the differential enthalpy of the adsorbed phase[18]. From the thermodynamic viewpoint, the heat of adsorption is determined using the phase information at specified temperature, pressure and concentration or the amount adsorbed. Despite the complexity involved in adsorption applications such as large database and combinations of adsorbent + adsorbate pairs, wide operating pressures from sub-atmospheric to supercritical, and temperatures spanning from cryogenic to several tens of Celsius; isosteric heat involved can be simply categorized into three natures with respect to the amount adsorbed or uptake. These types of isosteric heat of adsorption are: (1) constant value irrespective to the uptake, (2) decreasing with increasing uptake and (3) increasing with higher uptake[19]. Constant isosteric heat is usually associated with the adsorption on certain types of adsorbates onto an energetically homogenous adsorbent. For some pairs, isosteric heat reduces at higher uptake or loading due to the variability in energetic heterogeneity and weaker bonding at multi-layer adsorption whilst stronger lateral interactions between the adsorbed molecules at higher uptake contributes to the increase in the isosteric heat[19,20].

Despite for the observed variations in isosteric heat of adsorption, Clasius-Clayperon and Van't Hoff equations, hitherto, are commonly employed to estimate the isosteric heat for various adsorbent + adsorbate pairs[21–23]. Noted that these equations assume the adsorbed phase volume to be negligible as compared to the gaseous phase and it behaves as an ideal gas. Shen et al., reviewed different methods for the isosteric heat measurement and reported that the prediction using the Clasius-Clayperon type equation provides reasonably well agreement with the calorimetry experiments at sub-atmospheric pressures, yet they concluded that an average difference of about 2 kJ/mol remains to be sorted out even for low pressure adsorption processes[18]. Recently, Tian et al., reported the expression for the heat of adsorption using gas fugacity and validated with the generated "experimental data" for the differential heat of adsorption using grand canonical Monte Carlo simulations (GCMC)[26]. Attempts have been made to address the non-ideal gas behavior of the adsorbed phase[19,24,25], hitherto, very limited or no work has been carried out on the adsorbed phase volume, especially at low pressure adsorption scenarios. Adsorbed phase volume might be relatively small at low pressures whilst it would lead to substantial inaccuracies for high pressure adsorption. Thus, for accurate prediction of isotherm behaviors and isosteric heat of adsorption, a unified theoretical framework on adsorbed phase volume needs to be formulated. In this letter, we address the aforementioned issues and develop thermodynamic model where the contribution by the adsorbed phase volume is accounted for.

Generally, physisorption involves weak van der Waals forces between the adsorbed molecules and the surfaces of the adsorbent which is relatively easy to be reversed with the application of thermal energy (thermal swing) or lowering pressure (pressure swing). Figure 1 depicts the physical and thermodynamic insight of physisorption on a three-dimensional surface of an adsorbent material (silica gel surface), the intermolecular forces and the equilibrium chemical potential between the bulk and adsorbed phases.

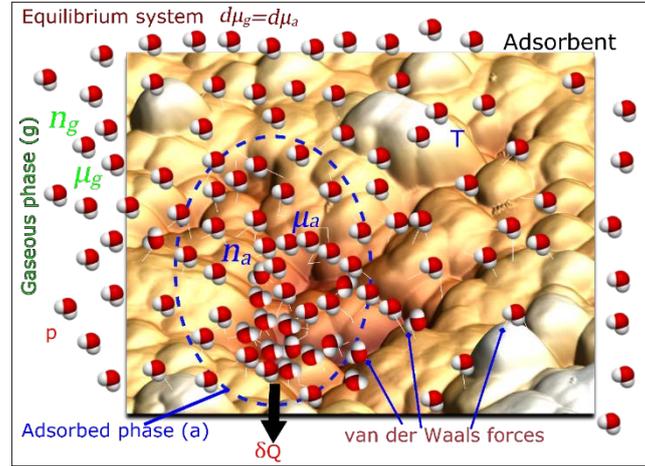

FIG. 1. Energetic and thermodynamic equilibrium involved in an adsorbent + adsorbate system (silica gel + water)

For an adsorbent + adsorbate system at equilibrium condition, the chemical potentials of the adsorbed phase and the gaseous bulk are equal i.e., $\mu_a = \mu_g$ and thus, $d\mu_a = d\mu_g$. The total differentials for the chemical potentials of the adsorbed and gaseous phases are written as,

$$d\mu_a = \left.\frac{\partial \mu_a}{\partial T}\right|_{x,p} dT + \left.\frac{\partial \mu_a}{\partial p}\right|_{x,T} dp + \left.\frac{\partial \mu_a}{\partial x}\right|_{p,T} dx \qquad (1)$$

$$d\mu_g = \left.\frac{\partial \mu_g}{\partial T}\right|_{p} dT + \left.\frac{\partial \mu_g}{\partial p}\right|_{T} dp. \qquad (2)$$

Invoking Gibbs-Duhem relation and equilibrium chemical potential between the phases, i.e., $d\mu_a = d\mu_g$, the following expression can be attained as,

$$-s_a dT + v_a dp + \left.\frac{\partial \mu_a}{\partial x}\right|_{p,T} dx = -s_g dT + v_g dp. \qquad (3)$$

Thus, the expression along the isostere i.e., constant uptake ($dx = 0$) can be given as,

$$\frac{s_g - s_a}{v_g - v_a} = \left.\frac{\partial p}{\partial T}\right|_x. \qquad (4)$$

Hence, the isosteric heat of adsorption equation, equivalent to the derivation by Hill[27], is expressed as,

$$\Delta h_{st} = T(v_g - v_a)\left.\frac{\partial p}{\partial T}\right|_x. \qquad (5)$$

Note here that if the adsorbed phase volume ($v_a$) is neglected with the ideal gas assumption, Eq. (5) simplifies to the well-known Clasius-Clayperon. Thus, one needs to obtain mathematical expressions for $v_a$ and the derivative $\left.\partial p/\partial T\right|_x$. The derivative term is to be calculated from a particular isotherm model where the adsorbed phase volume correction has to be incorporated. Invoking Dubinin-Astakhov (D-A) isotherm equation, the pressure gradient term becomes,

$$\left.\frac{\partial \ln p}{\partial T}\right|_x = \left.\frac{\partial \ln p_s}{\partial T}\right|_x + \frac{E}{RT^2}\left\{(-\ln\theta)^{\frac{1}{n}} + \frac{T}{nv_a}(-\ln\theta)^{\frac{1}{n}-1}\left.\frac{\partial v_a}{\partial T}\right|_x\right\}. \qquad (6)$$

Adsorbed phase volume, $v_a$, correction has been considered for adsorption at high pressure where volumetric thermal expansion coefficient is employed to estimate $v_a$ setting boiling or triple point as reference[28,29]. Akkimaradi et al. proposed the van der Waals volume[30] whilst Srinivasan et al. treated the adsorbed phase as another equilibrium phase satisfying the Gibbs equation and a linear function of temperature is assumed to relate the average densities of the adsorbed and bulk gas phases[31]. However, as stated by Hill, the properties of adsorbed phase with multi-layer adsorption will certainly depend on the pressure[27]. Thus, we reckon that the adsorbed phase is subjected to both the pressure and temperature and hence, a thermodynamically consistent expression for the adsorbed phase volume, $v_a$, is proposed as,

$$v_a = A \times T + B \times p^C \qquad (7)$$

Finally, the expression for the isosteric heat of adsorption for all temperature and pressure conditions accounting adsorbed phase volume is given by,

$$\Delta h_{st} = pT(v_g - v_a)\left[\left.\frac{\partial \ln p_s}{\partial T}\right|_x + \frac{E}{RT^2}\left\{(-\ln\theta)^{\frac{1}{n}} + \frac{T}{nv_a}(-\ln\theta)^{\frac{1}{n}-1}\left.\frac{\partial v_a}{\partial T}\right|_x\right\}\right] \qquad (8)$$

The proposed model is validated using five sets of published data from several research works where the isosteric heat of adsorption is experimentally measured using calorimetric method. We opined that it is essential to report the accuracy of any proposed model in predicting the isotherm data prior looking into the jumping correctness of the expression for isosteric heat of adsorption; since, it is always possible to reach excellent fitting accuracy for isosteric heat of adsorption yet the isotherm fittings might be notably poor. Figure 2 depicts the isotherm fitting for two types of adsorbent (NaX and silicalite) with three kinds of adsorbate ($CO_2$, $SF_6$, and $N_2$)[20,32,33] using the proposed adsorbed phase volume correction (Eq. 7 with Dubinin-Astakhov adsorption isotherm model) where excellent agreement with the experimental data is observed whilst the fitting accuracy is within ±3% for all sets of data.

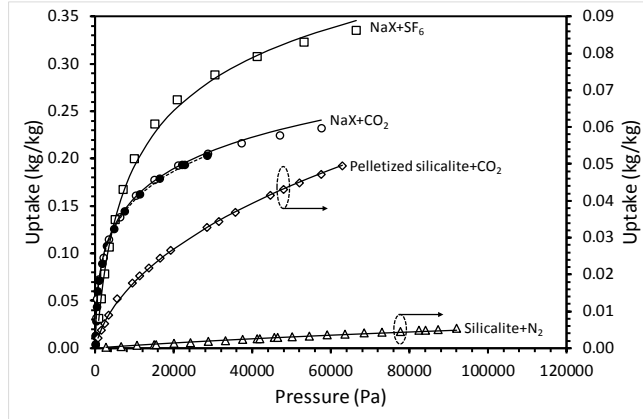

FIG. 2. Validation of the adsorption isotherm model using the proposed adsorbed phase volume correction (Eq. 7 and Dubinin-Astakhov adsorption isotherm model) for five sets of experimental data for the adsorption of: (□) $SF_6$ on Silicalite at 304.75K[32], (O) $CO_2$ on NaX at 304.55K[32], (◇) $CO_2$ on pelletized Silicalite at 305.15K[20] and (△) $N_2$ on Silicalite at 296.1K[33]

Figure 3 illustrates the prediction and validation of the isosteric heat of adsorption using Eq. (8) with the experimentally-measured data for the aforementioned adsorbent + adsorbate pairs. It is observed that the present model can accurately predict all three natures of isosteric heat of adsorption. Adsorption of carbon dioxide ($CO_2$) on NaX and silicalite exhibits decreasing trend with higher pressure (thus higher uptake) whilst the adsorption of nitrogen ($N_2$) on the silicalite demonstrates reasonably constant isosteric heat of adsorption irrespective of the uptake amount. Nevertheless, the adsorption of sulphur hexafluoride, ($SF_6$) on NaX shows increasing isosteric heat as adsorption commences to higher uptakes. Here, we observed that the predictions using Eq. (8) agree remarkably well with the calorimetrically-measured isosteric heat of adsorption for all sorption pairs.

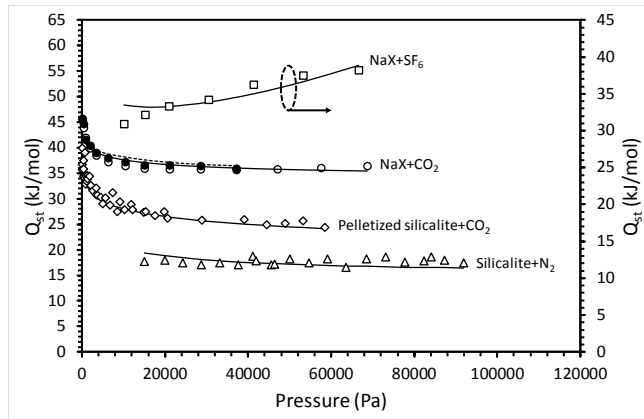

FIG. 3. Comparison and verification of the proposed model for the isosteric heat of adsorption (Eq. 8) with the experimentally (calorimetrically) measured data, the gradients for the pressure and adsorbed phase volume along the isostere are calculated using Dubinin-Astakhov isotherm model and Eq. 7 for the adsorption of: (□) $SF_6$ on Silicalite at 304.75K[32], (O) $CO_2$ on NaX at 304.55K[32], (◇) $CO_2$ on pelletized Silicalite at 305.15K[20] and (△) $N_2$ on Silicalite at 296.1K[33]

In summary, the theoretical development on the isosteric heat of adsorption is carried out based on the thermodynamic framework. The distinctive feature of the present work is that adsorbed phase volume is accounted for treating as a function of adsorption temperature and pressure, and the model is valid for adsorption from low to high pressures. The model is validated using calorimetrically-measured data and excellent prediction accuracy for both isotherm and the isosteric heat of adsorption is observed for different adsorbent + adsorbate pairs. The pressure gradient along an isostere is expressed as a function of the adsorbed phase volume. Hence, the present formulation can predict all types of isosteric heat of adsorption with superior accuracies as compared to conventionally used models such as Clasius-Clayperon and Van't Hoff equations.

This work has been funded by the Kyushu University Program for Leading Graduate School, Green Asia Education Center for their financial support to conduct this research.

**REFERENCES**

[1] D.G. Hartzog and S. Sircar, Adsorption **1**, 133 (1995).

[2] A.L. Myers, AIChE J. **48**, 145 (2002).

[3] S. Sircar, Ind. Eng. Chem. Res. **45**, 5435 (2006).

[4] J. Sun, T.D. Jarvi, L.F. Conopask, S. Satyapal, M.J. Rood, and M. Rostam-Abadi, Energy & Fuels **15**, 1241 (2001).

[5] N. Bimbo, V.P. Ting, A. Hruzewicz-Kolodziejczyk, and T.J. Mays, Faraday Discuss. **151**, 59 (2011).

[6] Y.. Aristov, G. Restuccia, G. Cacciola, and V.. Parmon, Appl. Therm. Eng. **22**, 191 (2002).

[7] A. Chakraborty, B.B. Saha, S. Koyama, and K.C. Ng, Appl. Phys. Lett. **90**, (2007).

[8] H.T. Chua, K.C. Ng, A. Malek, T. Kashiwagi, A. Akisawa, and B.B. Saha, Int. J. Refrig. **31**, 536 (2008).

[9] S.K. Henninger, H.A. Habib, and C. Janiak, J. Am. Chem. Soc. **131**, 2776 (2009).

[10] M.Z.I.Z.I. Khan, K.C.A.C.A. Alam, B.B.B. Saha, Y. Hamamoto, A. Akisawa, and T. Kashiwagi, Int. J. Therm. Sci. **45**, 511 (2006).

[11] A. Myat, K. Thu, Y.D. Kim, B.B. Saha, and K. Choon Ng, Energy **46**, 493 (2012).

[12] L.W. Wang, J.Y. Wu, R.Z. Wang, Y.X. Xu, S.G. Wang, and X.R. Li, (n.d.).

[13] A. Chakraborty, K.C. Leong, K. Thu, B.B. Saha, and K.C. Ng, Appl. Phys. Lett. **98**, 221910 (2011).

[14] J.W. Wu, M.J. Biggs, P. Pendleton, A. Badalyan, and E.J. Hu, Appl. Energy **98**, 190 (2012).

[15] S. Mitra, P. Kumar, K. Srinivasan, and P. Dutta, Int. J. Refrig. **58**, 186 (2015).

[16] S.M. Ali and A. Chakraborty, Appl. Therm. Eng. **90**, 54 (2015).

[17] K. Thu, Y.-D. Kim, M.W. Shahzad, J. Saththasivam, K.C. Ng, and K. Choon, in *Soc. Air-Conditioning Refrig. Eng. Korea, SAREK* (2014), pp. 469–477.

[18] D. Shen, M. Bülow, F. Siperstein, M. Engelhard, and A.L. Myers, Adsorption **6**, 275 (2000).

[19] A. Chakraborty, B.B. Saha, S. Koyama, and K.C. Ng, Appl. Phys. Lett. **89**, (2006).

[20] S. Sircar, R. Mohr, C. Ristic, and M.B. Rao, J. Phys. Chem. B **103**, 6539 (1999).

[21] B.J. Schindler and M.D. LeVan, Carbon N. Y. **46**, 644 (2008).

[22] K. Uddin, I.I.I.I.I. El-Sharkawy, T. Miyazaki, B.B.B.B. Saha, S. Koyama, H.S.H.-S.H.-S. Kil, J. Miyawaki, S.-H.S.-H.H. Yoon, B.B.B.B. Saha, S. Koyama, H.S.H.-S.H.-S. Kil, J. Miyawaki, and S.-H.S.-H.H. Yoon, Appl. Therm. Eng. **72**, 211 (2014).

[23] D.D. Do, *Adsorption Analysis: Equilibria and Kinetics* (Imperial College Press, 1998).

[24] A. Chakraborty, B.B. Saha, K.C. Ng, S. Koyama, and K. Srinivasan, Langmuir **25**, 2204 (2009).

[25] A. Chakraborty, B.B. Saha, I.I. El-Sharkawy, S. Koyama, K. Srinivasan, and K.C. Ng, High Temp. - High Press. **37**, 109 (2008).

[26] Y. Tian and J. Wu, Langmuir **33**, 996 (2017).

[27] T.L. Hill, J. Chem. Phys. **17**, 520 (1949).

[28] B.B. Saha, S. Koyama, I.I. El-Sharkawy, K. Habib, K. Srinivasan, and P. Dutta, J. Chem. Eng. Data **52**, 2419 (2007).

[29] S. Ozawa, S. Kusumi, and Y. Ogino, J. Colloid Interface Sci. **56**, 83 (1976).

[30] B.S. Akkimaradi, M. Prasad, P. Dutta, and K. Srinivasan, Carbon N. Y. **40**, 2855 (2002).

[31] K. Srinivasan, B.B. Saha, K.C. Ng, P. Dutta, and M. Prasad, Phys. Chem. Chem. Phys. **13**, 12559 (2011).

[32] J.A. Dunne, R. Mariwala, M. Rao, S. Sircar, R.J. Gorte, and A.L. Myers, Langmuir **12**, 5896 (1996).

[33] J.A. Dunne, R. Mariwala, M. Rao, S. Sircar, R.J. Gorte, and A.L. Myers, Langmuir **12**, 5888 (1996).